
\def\slD{\raise.15ex\hbox{$/$}\kern-.57em\hbox{$D$}}
\def\cA{{\cal A}}
\def\cD{{\cal D}}
\def\cH{{\cal H}}
\def\cL{{\cal L}}
\def\cR{{\cal R}}
\def\cS{{\cal S}}
\def\cZ{{\cal Z}}
\def\pr{{\partial \over \partial r}}
\def\prr{{\partial^2 \over \partial r^2}}
\def\pt{{\partial \over \partial \theta}}
\def\ar{A_{r}}
\def\at{A_{\theta}}
\def\att{a_{\theta}}
\def\a{\hat \cA}
\def\aR{\hat \cA |_{R}}
\def\Gr{\slD[A']^{-1}_{(x,y)}}
\def\erfc{{\it erfc}}
\def\if{{\it if}}
\def\j{{\bf j}}
\def\ie{{\it i.e.,\ }}
\def\etc{{\it etc.\ }}

\rightline{August 24, 1992}

\title FUNCTIONAL DETERMINANT AND GREEN FUNCTIONS FOR A FERMIONIC GAUGE
THEORY ON THE DISK

\author Enrique F. Moreno

\affil Department of Physics
       University of Illinois at Urbana-Champaign
       1110 W. Green St., Urbana, IL 61801, USA.

\abstract {We study a theory of Dirac fermions on a disk in presence of an
electromagnetic field. Using the heat-kernel technique we compute the
functional determinant which results after decoupling the zero-flux gauge
degrees of freedom from the fermions. We also compute the Green functions of
the remaining fermionic theory with the appropriate boundary conditions.
Finally we analyze the coset model associated to this gauge theory and compute
all its correlations functions.}
\endtopmatter

\head{\bf SECTION I}

An interesting problem in Quantum Field Theory is the modification of the
physical quantities due to the presence of boundaries. There are many important
reasons which justifies this statement: the boundary effects on the chiral
anomalies, as in the chiral bag model, the study of open strings, the
connection of QFT on manifolds with boundary with statistical mechanics
systems, \etc.

For a fermionic theory defined on a manifold with boundary it is known that
only special boundary conditions imposed to the fermions (spectral boundary
conditions) are consistent with charge conjugation and chiral
invariance\refto{horda}. Also the usual index theorem for compact manifolds is
no longer valid. Instead the Atiyah-Patodi-Singer\refto{aps} (APS) theorem
holds. This theorem relates the index of the Dirac operator with the global
anomaly and a {\it new} surface  contribution proportional to the spectral
asymmetry of the Dirac operator restricted to the boundary.

Due to the existence of a non-zero magnetic flux $\Phi$, the Dirac operator has
$[\Phi]$ non-trivial zero modes which makes the definition of the fermionic
determinant ill defined. However, for two gauge fields having the same total
magnetic flux, the ratio of determinants can be defined without divergencies
(see \ref{solo}).
One of the results of this work is the computation, for the two-dimensional
case, of the fermionic functional determinant.  As expected by the work of APS,
the determinant acquires a contribution supported at the boundary which takes
into account the edge degrees of freedom. An interesting point is that the
effective action for the gauge field can be written as a Liouville type action,
with a conveniently chosen background gravitational field, including not only a
coupling with the Gaussian curvature  but also a coupling with the geodesic
curvature of the boundary. In this way the APS index theorem takes he form of
the Gauss-Bonnet theorem.

Using this result we can reduce the problem of fermions interacting with an
arbitrary gauge field to the case in which the gauge field is
independent of the angular variable. The advantage of this
procedure is that for the theory of fermions interacting with a
$\theta$-independent gauge field all the Green functions can be easily
integrated. We compute this Green functions with the appropriate boundary
conditions and in particular we recover the relation between the chiral
anomaly and the index theorem in an ``operational" way.

Motivated by the study of the coset models we consider the case when the gauge
field is also a dynamical variable. In this case, besides  the fermionic
contractions, the correlation functions also contains bosonic vertex operators
correlators, governed by the  Liouville action mentioned before. The fermionic
zero modes of the theory forces the total charge be equal to $[\Phi]$, and the
effect of the geometry is manifested in the appearance of an  ``image" vertex
operator for each vertex operator in the correlation function.
For this reason the conformal properties of the fields (critical dimensions) in
the neighborhood of the boundary are different of the values in the
bulk\refto{cardy}.

The organization of the paper is as follows: In section II we present our
problem and state the role of the boundary conditions.
In section III we compute the functional determinant. We show that the
effective theory for the gauge field can be written as a Liouville
theory with a conveniently chosen background metric. The boundary contribution
appears as an interaction with the geodesic curvature of the boundary. In
section IV we compute the fermionic Green functions. Also we compute the
modification of the chiral anomaly and its relation with the index theorem in
this geometry. Finally in section V we study the coset theory which appears
when we integrate the gauge field. We compute all the correlation
functions.

\head{\bf SECTION II}

Consider the following Dirac Lagrangian
$$
\cL ={1\over 4\pi} {\bar \psi} i \slD \psi~~,~~~~~i\slD = \gamma^{\mu}\left(
i\partial_{\mu} + A_{\mu}\right) \eqno(2.1)
$$
defined on the disc ${\bf D}=\{ {\vec x} \in {\cR}^2 : |{\vec x}| \leq R
\}$. $A_{\mu}$ is a background gauge field which on the boundary takes the
values
$$
\at(R,\theta)=\aR.\eqno(2.2)
$$
The Lagrangian (2.1) is gauge invariant and we choose the radial gauge
$\ar(r,\theta)=0$. In this gauge, in polar coordinates,
the Dirac operator takes the form
$$
i \slD =\pmatrix{0&D_{z}\cr D_{\bar z}&0\cr}\eqno (2.3)
$$
$$
D_{z}=e^{i\theta}\left(-{\partial \over \partial r} + {i\over r}{\partial
\over \partial \theta}+ {A_{\theta} \over r}\right)\eqno(2.4 a)
$$
$$
D_{\bar z}=e^{-i\theta}\left({\partial \over \partial r} + {i\over
r}{\partial  \over \partial \theta}+ {A_{\theta} \over r}\right).\eqno(2.4 b)
$$
The magnetic flux
through the disc is determined by the boundary condition (2.2)
$$
\Phi={1\over 2\pi}\int dr d\theta \pr \at = {1\over 2\pi}\oint d\theta \aR
(\theta).\eqno(2.5)
$$

The quantum theory defined by the Lagrangian (2.1), in presence of an {\it
arbitrary} gauge field can be reduced, after an appropriate chiral rotation, to
the same theory in presence of a $\theta$ {\it independent} gauge field \ie
$\pt \at =0$ (the advantage of this change is that in the later case all the
fermionic Green functions can be exactly integrated). The procedure is the
following.  The generating functional of correlation functions is given, in
euclidean space, by
$$
\cZ [{\bar \zeta},\zeta; A] = \int D{\bar \psi} D\psi e^{-S[{\bar \psi},\psi;
A] + \int ({\bar \zeta} \psi + {\bar \psi}\zeta )}\eqno(2.6)
$$
If $\at '(r)$ is a $\theta$
independent gauge field satisfying
$$
\Phi[\at '] = \Phi[\at]~~~~~~or\eqno(2.7)
$$
$$
\at '(R) = {1\over 2\pi} \oint d\theta \at (R,\theta).\eqno(2.8)
$$
we can decompose the gauge field $\at$ as
$$
\at (r,\theta) = \at ' (r) + \att (r,\theta)\eqno(2.9)
$$
with $\att (r,\theta)$ a zero flux field. Writing
$$
\att (r,\theta) = \pt \omega (r,\theta) -r\pr \eta (r,\theta)\eqno(2.10)
$$
with the condition
$$
a_{r}(r,\theta) = \pr \omega (r,\theta) + {1\over r}\pt \eta (r,\theta)=0
\eqno(2.11)
$$
and performing the fermionic change of variables
$$
\psi \to e^{i\omega + \gamma^5 \eta} \psi\eqno(2.12 a)
$$
$$
{\bar \psi} \to  {\bar \psi} e^{-i\omega + \gamma^5 \eta}\eqno(2.12 b)
$$
the $\att$ field decouple from the fermions
$$
{\bar \psi} \gamma^{\mu}\left( i\partial_{\mu} + A'_{\mu} + a_{\mu}\right)
\psi \to  {\bar \psi} \gamma^{\mu}\left( i\partial_{\mu} + A'_{\mu}\right)
\psi. \eqno(2.13)
$$

This chiral change of variables has a non-trivial jacobian proportional to the
ratio of the Dirac determinants
$$
J[A] = C(A)~ {det [\gamma^{\mu}\left( i\partial_{\mu} + A'_{\mu} +
a_{\mu}\right)] \over det [\gamma^{\mu}\left( i\partial_{\mu} +
A'_{\mu}\right)]}\eqno(2.14)
$$
where the factor $C(A)$ comes from the sources term. Note that although the
operators $\slD [A'+a]$ and $\slD [A']$ have zero modes and then their
determinants are ill defined, the ratio is well defined\refto{solo}  because
condition (2.6) ensures that the number of zero modes of both operators is the
same. Finally after a little of algebra we can prove that the generating
functional takes the form\refto{brazil}
$$
\cZ [{\bar \zeta},\zeta; A] = J[A] e^{\int {\bar \zeta} e^{i\omega +
\gamma^5 \eta} (\slD [A'])^{-1} e^{-i\omega + \gamma^5 \eta} \zeta}\times
$$
$$
det'\slD [A'] \prod_{zero~modes}(\int \varphi^{\dagger}_{0 n} e^{-i\omega +
\gamma^5 \eta}  \zeta) (\int {\bar \zeta} e^{i\omega + \gamma^5 \eta}
\varphi_{0 n}) \eqno(2.15)
$$
where $(\slD[A'])^{-1}$ is the Green function of the operator $\slD[A']$
satisfying
$$
\slD [A']_{x} \slD[A']^{-1}_{(x,y)} = \delta (x,y) - \sum_{zero~modes}
\varphi_{0 n} (x)~\varphi^{\dagger}_{0 n}(y),\eqno(2.16)
$$
$det'\slD$ is the (regularized) product of non-zero eigenvalues of $\slD$ and
$\{\varphi_{0 n}\}$ is a set of orthonormal zero modes of $\slD[A']$.

Before computing the determinant we have to talk about the boundary conditions.
The quantum problem defined by the action in (2.1) is incomplete until we
specify the proper boundary conditions ensuring the  self-adjointness of the
Dirac operator. It is well known that for this problem local boundary
conditions (Dirichlet, Neumann) are incompatible with the chiral and charge
conjugation invariance or, from a mathematical point of view, there exists
topological obstructions which prevents the use of local boundary conditions in
this kind of operators. All this problems are solved with the so-called {\it
spectral boundary conditions} which are defined over the eigenspaces of the
differential operator obtained by restricting the Dirac operator to the
boundary.
In order to present these issues more clear let us define the operators
$$
\cD=-\pr + \a~~,~~~~\cD^{\dagger}=\pr + \a\eqno(2.17)
$$
$$
\a= {i\over r}\pt +{\at \over r}\eqno(2.18)
$$
related with $D_{z}$ and $D_{\bar z}$ by
$$
D_{z}={e^{-i\theta/2} \over \sqrt{r}} \cD e^{-i\theta/2} \sqrt{r}\eqno(2.19)
$$
$$
D_{\bar z}={e^{i\theta/2} \over \sqrt{r}} \cD^{\dagger} e^{i\theta/2} \sqrt{r}
\eqno(2.20)
$$
and a new spinor field $\chi$
$$
\chi=\sqrt{r} e^{-i\theta/2 \gamma^{5}}\psi.\eqno(2.21)
$$

In terms of this new operators the action takes the form
$$
S={1\over 4\pi}\int dr d\theta \left({\bar \chi}_1 \cD \chi_2 + {\bar \chi}_2
\cD^{\dagger} \chi_1\right)\eqno(2.22)
$$
and the natural inner product between spinors is
$$
\left(\chi,\varphi\right)=\int dr d\theta \left(\chi_{1}^{*} \varphi_{2} +
\chi_{2}^{*} \varphi_{1}\right).\eqno(2.23)
$$
By integrating by parts we easily find that the condition for the hermiticity
of $i\slD$ is
$$
\oint  d\theta \chi_{1}^{*}(R,\theta) \varphi_{2}(R,\theta)=0
{}~~~~~~~for~all~~\chi_{1}~~and~~\varphi_{2}.\eqno(2.24)
$$

The spectral boundary conditions are simply conditions over the expansion of
the spinor components at the boundary in terms of the eigenfunctions
$\nu_{\lambda}(\theta)$ of $\aR$ ($\aR={i\over R}\pt +{\at (R) \over R}$).
Examples of them are

$$
\chi_{1}(R,\theta)=\sum_{\lambda > \alpha} \chi_{1}^{\lambda} \nu_{\lambda}
(\theta)\eqno(2.25 a)
$$
$$
\chi_{2}(R,\theta)=\sum_{\lambda \leq \alpha} \chi_{2}^{\lambda} \nu_{\lambda}
(\theta)\eqno(2.25 b)
$$
or
$$
\chi_{1}(R,\theta)=\sum_{\lambda \geq \beta} \chi_{1}^{\lambda} \nu_{\lambda}
(\theta)\eqno(2.26 a)
$$
$$
\chi_{2}(R,\theta)=\sum_{\lambda < \beta} \chi_{2}^{\lambda} \nu_{\lambda}
(\theta)\eqno(2.26 b)
$$
labeled by the arbitrary parameter $\alpha$ (or $\beta$). Clearly the equations
(2.25) (or (2.26)) solves the hermiticity condition (2.24).
With this boundary conditions the Dirac operator has an index governed by the
Atiyah-Patodi-Singer (APS) index formula\refto{aps} which for our special case
takes the form\refto{horda, niemi, rusos}
$$
index \slD = n_1 - n_2 = \Phi - \left( \Phi - int_{-}(\Phi + {1\over 2} +
\alpha)\right)\eqno(2.27)
$$
for the b.c. (2.25) or
$$
index \slD = n_1 - n_2 = \Phi - \left( \Phi - int_{+}(\Phi - {1\over 2} +
\beta)\right)\eqno(2.28)
$$
for the b.c. (2.26). Here $int_{-}(x)$ ($int_{+}(x)$) is the first integer less
or equal (greater or equal) than $x$. In both expressions (2.27) and (2.28) the
first term of the r.h.s comes from the usual integral of the anomaly density
and the second term comes from boundary contributions.
Here, as in the case of two dimensional boundaryless manifolds, holds a
{\it vanishing theorem}: either all the zero modes are right handed or all are
left handed.

Even when any election of the spectral parameter $\alpha$ (or $\beta$) gives a
good result let us choose a particular one. To fix this value we use the
natural argument\refto{rusos} that the set of zero modes could be extended to a
set of square-integrable zero modes when we extend the domain of definition of
the fields from the disk to the whole plane ${\cR}^2$. By a direct computation
of the zero modes (see also the index formulas (2.27) and (2.28)) we can verify
that this condition is reached when\refto{rusos}
$$
\eqalign{&\alpha=-{1\over 2}~~~~~if~\Phi>0~~~~~~~or\cr
&\beta={1\over
2}~~~~~if~\Phi<0.\cr}\eqno(2.29)
$$
We then choose the boundary conditions defined by the values (2.19) and we
write in a short form
$$
index \slD = [\Phi] = \cases {int_{-}(\Phi)&if $\Phi >0$\cr int_{+}(\Phi)
&if $\Phi <0$\cr}.\eqno(2.30)
$$

\head{\bf SECTION III}

In equation (2.15) we could write the generating functional {\cZ} for an
arbitrary gauge field $A$ in terms of the generating functional of a
$\theta$-independent gauge field $A'$ times a jacobian proportional to the
ratio of the determinants
$$
J[A] = C(A)~ {det (\gamma^{\mu}\left( i\partial_{\mu} + A'_{\mu} +
a_{\mu}\right)) \over det (\gamma^{\mu}\left( i\partial_{\mu} +
A'_{\mu}\right))}\eqno(3.1)
$$
where $C(A)$ is a factor originated in the Grassmann sources.
Using well known techniques for the computation of determinants of singular
operators\refto{solo} we find that the ratio in equation (3.1) can be written
as
$$
{det~i\slD [A'_{\mu}+a_{\mu}]\over det~i\slD [A'_{\mu}]}=e^{-\int_{0}^{1}
\omega '(t)~dt}\eqno(3.2)
$$
where
$$
\omega '(t)=2 \int_{\bf D} d^2x~tr K_2 [i\slD_{t};{\vec x},{\vec
x}]~\gamma^5 \eta ({\vec x})+
$$
$$
-2\sum_{zero~modes} \int_{\bf D} d^2x~tr\left(\varphi_{0 n} (t;{\vec
x})~\gamma^5 \varphi^{\dagger}_{0 n}(t;{\vec x})\right)\eta({\vec
x}).\eqno(3.3)
$$
Here $i\slD_{t}=i\slD[A' + (1-t)a]$, $K_2 [i\slD;{\vec x},{\vec
x}]$ is the constant term in the analytic expansion of $<x|e^{-s\slD \slD}|x>$,
$\eta({\vec x})$ is the field defined in
equation (2.9) and $\{ \varphi_{0 n} (t;x) \}$ is an orthogonal set of zero
modes of $i\slD_{t}$.

As it was pointed out in \Ref{brazil} the contribution to $J[A]$ from
the second term in the r.h.s. of equation (3.3) cancels exactly the factor
$C(A)$ in the jacobian coming from the Grassmann sources. Hence the value of
the jacobian is determined only by the kernel $K_2 [i\slD _{t}]$ which can be
computed following APS\refto{aps} separating the contributions of the boundary
and of the bulk.
Indeed we decompose the disk {\bf D} into an inner disk ${\bf D}_{R-\epsilon}$
of radius $R-\epsilon$ and the (infinitesimal) exterior ring ${\bf A}=\{ {\vec
x}\in {\cR}^2 : R-\epsilon < |{\vec x}| \leq R\}$. In the inner disk we can
evaluate $K_2$ in a conventional way and we obtain for $\omega'(t)$ the usual
value for closed compact manifolds\refto{determinantes}
$$
{1\over 2\pi}\int_{{\bf D}_{R-\epsilon}}d^2x~\{F_{x y}[A']+(1-t)F_{x y}[a]
\}~\eta({\vec x}).\eqno(3.4)
$$

The contribution of the boundary (exterior ring) to the jacobian is given by
the regular term in the limit
$$
2\displaystyle\lim_{s\to0}~{\int_{R-\epsilon}^{R} rdr~\oint d\theta~<r|\otimes
<\theta| e^{s D_{z}D_{\bar z}} - e^{s D_{\bar z}D_{z}}
|r>\otimes|\theta>~\eta(r,\theta)}\eqno(3.5)
$$
which can be notably simplified if we note that we can replace $D_{z}$ by $\cD$
and $D_{\bar z}$ by $\cD^{\dagger}$ in the trace of the kernel operator and
also, near the boundary,  we can approximate  the operators $\cD$ and
$\cD^{\dagger}$ by $-\pr + \aR$ and $\pr + \aR$ respectively. Hence in the
eigenspace of eigenvalue $\lambda$ of the operator  $\aR$ we have
$$
<r|e^{s D_{z}D_{\bar z}}|r>\approx <r|e^{s (-\prr + \lambda^2)}|r>\eqno(3.6
a)
$$
$$
<r|e^{s D_{\bar z}D_{z}}|r>\approx <r|e^{s (-\prr + \lambda^2)}|r>.\eqno(3.6
b)
$$
The boundary conditions of the operator $-\prr + \lambda^2$ are inherited from
the ones of
the operators $\cD$ and $\cD^{\dagger}$ (equations (2.25) and (2.26)) and reads
$$
\chi_{1}(R,\theta)=0~~for~~\lambda\leq -1/2~~ (or~~\lambda < 1/2)
$$
$$
(\pr + \lambda)\chi_{1}(R,\theta)=0~~for~~\lambda > -1/2~~ (or~\lambda \geq
1/2)\eqno(3.7)
$$
for the case $(3.6 a)$ and
$$
\chi_{2}(R,\theta)=0~for~\lambda > -1/2~~(or~~\lambda \geq 1/2)
$$
$$
(\pr + \lambda)\chi_{2}(R,\theta)=0~~for~~\lambda \leq -1/2~~(or~\lambda <
1/2)\eqno(3.8)
$$
for the case $(3.6 b)$.

The heat kernel in (3.6) can be computed by making a Laplace transform in the
variable $s$ in the heat equation
$$
\left(\partial_{s} -\prr + \lambda^2 \right) K(r,r',s)=0\eqno(3.9)
$$
with the boundary condition
$$
\displaystyle\lim_{s\to0}{K(r,r',s)}\to \delta(r-r')\eqno(3.10)
$$
and the boundary conditions defined by equations (3.7) and (3.8). After summing
over the whole spectrum of $\aR$ we get for the trace of the heat kernel
$$
<r| e^{s D_{z}D_{\bar z}} - e^{s D_{\bar z}D_{z}}|r> = {1\over 2\pi}
\sum_{\lambda > -1/2 ~(\lambda \geq 1/2)} {1\over 2} {d\over dr}\{e^{2\lambda
(R-r)} \erfc\left( {R-r\over \sqrt{s}} + \lambda \sqrt{s}\right) \} +
$$
$$
-{1\over 2\pi} \sum_{\lambda \leq -1/2 ~(\lambda < 1/2)} {1\over 2} {d\over
dr}\{e^{-2\lambda (R-r)} \erfc\left( {R-r\over \sqrt{s}} - \lambda
\sqrt{s}\right) \}\eqno(3.11)
$$
where $\erfc(x)={2\over \sqrt{\pi}}\int_{x}^{\infty} e^{-t^2}dt$ is the
complementary error
function. The eigenvalues of the operator $\aR$ are
$$
\lambda = {1\over R} (n+{1\over 2}+\Phi)\eqno(3.12)
$$
and the eigenvectors
$$
\nu_{\lambda}(\theta)={1\over \sqrt{2\pi}} e^{i \int_{0}^{\theta} d\theta
(\at - R\lambda)}\eqno(3.13)
$$
have constant square modulus.

It can be inferred from equation (3.11) that when $s\to 0$ only the superior
limit $r=R$ contributes in the integral in (3.5). Moreover we can prove that at
order less than $\sqrt{s}$ the integral in (3.5) takes the form
$$
-{1\over 2\pi}\oint \eta(R,\theta) d\theta~\left(\sum_{\lambda\leq -1/2}
\erfc(\lambda\sqrt{s}) - \sum_{\lambda > -1/2} \erfc(-\lambda\sqrt{s})\right) +
o(\sqrt{s}).\eqno(3.14)
$$

In the limit $s\to 0$, the expression inside the brackets in equation (3.14) is
related with the  spectral asymmetry of the operator $\aR$. In fact, a naive
computation of this limit leads to the (ill defined) quantity
$$
{1\over 2\pi}\oint \eta(R,\theta) d\theta~\left(\sum_{\lambda\leq -1/2}
1 - \sum_{\lambda > -1/2} 1 \right)\eqno(3.15)
$$
which counts the excess of eigenvalues equal or less than $-1/2$ over the
eigenvalues greater than $-1/2$.
Of course this limit is finite if is taken carefully. For example using the
identity\refto{mike,rusos}
$$
\sum_{j=0}^{\infty} \erfc((j+\rho)t) -  \sum_{j=1}^{\infty} \erfc((j-\rho)t) =
1-2\rho + o(t)~~~~(\rho\in [0,1])\eqno(3.16)
$$
we can show that
$$
\displaystyle {1\over 2}\lim_{s\to 0} {\left(\sum_{\lambda\leq -1/2}
\erfc(\lambda\sqrt{s}) - \sum_{\lambda > -1/2} \erfc(-\lambda\sqrt{s})\right)}
=\Phi - [\Phi].\eqno(3.17)
$$

With this result for the boundary contribution to the jacobian, the complete
jacobian (3.1) takes the form
$$
J=exp\left( {1\over 2\pi}\int \eta \triangle \eta d^2 x - {1\over \pi}\int
\eta F_{x y}[A'] d^2 x + (\Phi - [\Phi]){1\over \pi}\oint \eta(R,\theta)
d\theta \right).\eqno(3.18)
$$

We can write this last equation in a more elegant form if we use the analogy
with the Liouville action\refto{yo}.
Writing the field $A'$ as
$$
A'_{\mu}=\partial_{\mu}\alpha - \epsilon_{\mu
\nu}\partial_{\nu}\omega\eqno(3.19)
$$
the electromagnetic stress tensor takes the form
$$
F_{x y}[A']=\triangle \omega\eqno(3.20)
$$
(we use the freedom in the field $\alpha$ to fix the gauge  $A'_{r}=0$) and
satisfies
$$
{1\over \pi}\int_{\bf D} \triangle\omega d^2 x = {R\over 2\pi}\oint d\theta
\pr\omega|_{r=R}  = \Phi.\eqno(3.21)
$$

Now we introduce on the disk the conformal metric
$$
g_{\mu \nu} = e^{-{2\over [\Phi]}\omega}\delta_{\mu \nu}.
\eqno(3.22)
$$
The Gaussian curvature associated with this metric is given by
$$
\cR={1\over [\Phi]} e^{{2\over [\Phi]}\omega} \triangle \omega\eqno(3.22)
$$
and the geodesic curvature of the boundary by
$$
k_{g}=e^{{1\over [\Phi]}\omega} \left({1\over R} - {1\over [\Phi]}\pr
\omega|_{r=R}\right).\eqno(3.23)
$$
Also the differential elements of volume and arc are given, respectively, by
$$
dv= e^{-{2\over [\Phi]}\omega} d^{2}x ~~~and~~~ds=e^{-{1\over [\Phi]}\omega}
(dr
+ rd\theta).\eqno(3.24)
$$

Because $\omega|_r$ is independent of $\theta$ (see section II) we have
$$
[\Phi]\oint_{\partial {\bf D}} k_{g} \eta ds = \left([\Phi] - \Phi\right)\oint
\eta(R,\theta) d\theta\eqno(3.25)
$$
and we can rewrite the jacobian in (3.18) as a bosonic field interacting with a
gravitational background field
$$
J=exp\left({1\over 2\pi}\int_{\bf D} \eta \triangle \eta \sqrt{g}d^2 x -
{[\Phi] \over \pi}\left(\int_{\bf D}  \cR \eta \sqrt{g}d^2 x +\oint_{\partial
{\bf D}} k_{g} \eta ds\right)\right). \eqno(3.26)
$$

In this expression the index formula, equation (2.30), is a consequence of the
Gauss-Bonnet theorem
$$
\int_{\bf D} \cR \sqrt{g}d^2 x + \oint_{\partial {\bf D}}
k_{g} ds= 2\pi.\eqno(3.27)
$$
For a constant field $\eta$ the ratio of determinants
in (3.2) is equal to $1$ and we infer from equation (3.3) that the logarithm of
the jacobian is proportional to the index of $i\slD$. Then
$$
e^{-\eta~2index~i\slD}=J[\eta=const.]=e^{-\eta~2[\Phi]{1\over 2\pi}(\int \cR
dv + \oint k_{g} ds)}\eqno(3.28)
$$
and using equation (3.27) we recover the index formula (2.30).

\head{\bf SECTION IV}

We have just computed in the previous section, the chiral jacobian related to
the change of variables (2.12 a) and (2.12 b). Then the only ingredient we need
in order to recover the generating functional of correlation functions is to
compute the fermionic propagator (2.16) with the appropriate boundary
conditions defined in equations (2.25) and (2.26).  This Green function can be
computed explicitly in virtue of the $\theta$-independence of the gauge field
$\at '(r)$. Following the procedure of section II we define the Green function
$\cH({\vec x}, {\vec y})$ related with $\Gr$ by
$$
\slD[A']^{-1}_{(x,x')} = {1\over \sqrt{r}} e^{-i \gamma^5 \theta/2}
\cH(r,\theta;r',\theta ')
{1\over \sqrt{r'}} e^{i \gamma^5 \theta'/2}\eqno(4.1)
$$
which satisfies the equation
$$
\pmatrix{0&\cD\cr \cD^{\dagger}&0\cr}~\cH(r,\theta;r',\theta ') =
\delta(r-r')\delta(\theta-\theta') - \sum_{zero~modes}
\chi^{0 n} (r,\theta)~\chi^{0 n \dagger}(r',\theta').\eqno (4.2)
$$
(The $\chi^{0 n}$ are related with the $\varphi_{0 n}$ of equation (2.16) by
the relation (2.21)).
Writing the matrix $\cH(x,y)$ in components
$$
\cH=\pmatrix{0&\cH_{+}\cr \cH_{-}&0\cr}\eqno(4.3)
$$
we have the following equations for $\cH_{+}$ and $\cH_{-}$
$$
\cD~\cH_{-}(r,\theta;r',\theta')=\delta(r-r')\delta(\theta-\theta')-
\sum_{zero~modes} \chi^{0 n}_{1} (r,\theta)~\chi^{0 n *}_{1}(r',\theta')
\eqno (4.4)
$$
$$
\cD^{\dagger}~\cH_{+}(r,\theta;r',\theta')=\delta(r-r')\delta(\theta-\theta')
- \sum_{zero~modes} \chi^{0 n}_{2} (r,\theta)~\chi^{0 n *}_{2}(r',\theta')
.\eqno (4.5)
$$
Note that, modulo the zero modes contribution, equation (2.22) says that
$$
\cH_{-}(r,\theta;r',\theta')= -{1\over 4\pi}<\chi_{2}(r,\theta)~{\bar
\chi}_{1}(r',\theta')>\eqno(4.6 a)
$$
$$
\cH_{+}(r,\theta;r',\theta')= -{1\over 4\pi}<\chi_{1}(r,\theta)~{\bar
\chi}_{2}(r',\theta')>.\eqno(4.6 b)
$$

As we mentioned in section II all the zero modes are of the same chirality,
positive if $\Phi>0$ or negative if $\Phi<0$. For definiteness we will consider
the case $\Phi>0$ in which there are $[\Phi]$ zero modes $\chi_{1}^{0}$
$$
\chi_{1 n}^{0}={C_n\over \sqrt{2\pi}}e^{-\int_0^r d\rho \at(\rho)/\rho}
r^{n-1/2}e^{i(n-1/2)\theta},~~~~n=1,...,[\Phi]\eqno(4.7)
$$
and none $\chi_{2}^{0}$.

As usual we expand the Green function $\cH$ in eigenfunctions of the operator
$\aR$. Thus we have, in the orthogonal complement to the kernel of
$\cD^{\dagger}$
$$
\cH_{-}|_{\bot ker\cD^{\dagger}}(r,\theta;r',\theta')=\sum_{\lambda}
g_{\lambda}(r,r') \nu_{\lambda}(\theta)\nu_{\lambda}^{*}(\theta')\eqno(4.8)
$$
where
$$
\nu_{\lambda}(\theta)={1\over 2\pi} e^{-i(n+1/2)\theta}~,~~~~\lambda={1\over
R}(n+{1\over 2}+\Phi)\eqno(4.9)
$$
are the eigenfunctions and eigenvalues of $\aR$.
The functions $g_{\lambda}$ satisfy the differential equations
$$
-\pr g_{\lambda}(r,r') + {1\over r}( n+{1\over 2}+\at(r))g_{\lambda}(r,r')=
\delta(r-r')\eqno(4.10)
$$
with the boundary conditions
$$
g_{\lambda}(R,r')=0~~~~~~\if~R\lambda>-{1\over 2}~~~(n+1>-\Phi)\eqno(4.11)
$$
$$
g_{\lambda}(r,R)=0~~~~~~\if~R\lambda\leq -{1\over
2}~~~(n+1\leq-\Phi).\eqno(4.12)
$$
The solution is straightforward and we get
$$
g_{\lambda}(r,r')=-{1\over 2}e^{\int_{r'}^{r} d\rho \at(\rho)/\rho}
\left({r\over r'}\right)^{n+1/2}\times \cases {sign(r-r')-1&if $\lambda >-1/2$
\cr sign(r-r')+1&if $\lambda \leq-1/2$.\cr}\eqno(4.13)
$$
Inserting this solutions in equation (4.8) we have
$$
\cH_{-}|_{\bot ker\cD^{\dagger}}(r,\theta;r',\theta')=
$$
$$
=e^{\int_{r'}^{r}
d\rho \at(\rho)/\rho} \times \{\sum_{n+1\leq -\Phi} {1\over
2}(1-sign(r-r'))\left({r\over r'}\right)
^{n+1/2} \nu_{\lambda}(\theta)\nu_{\lambda}^{*}(\theta')+
$$
$$
- \sum_{n+1>  -\Phi}  {1\over 2}(1+sign(r-r'))\left({r\over r'}\right)^{n+1/2}
\nu_{\lambda}(\theta) \nu_{\lambda}^{*}(\theta')\}\eqno(4.14)
$$
and using the explicit form of the functions $\nu_{\lambda}$ (equation (4.9))
$$
\cH_{-}|_{\bot ker\cD^{\dagger}}(r,\theta;r',\theta')=-{1\over 2\pi}
e^{\int_{r'}^{r}
d\rho \at(\rho)/\rho}~\left({{\bar \omega}\over {\bar z}}\right)^{[\Phi]}
{\sqrt{{\bar z}{\bar \omega}}\over {\bar z} - {\bar \omega}}\eqno(4.15)
$$
where ${\bar z}=re^{-i\theta}$ and ${\bar \omega}=r'e^{-i\theta'}$.

The contribution to the Green function coming from the kernel of
$\cD^{\dagger}$ (which gives the contribution of the zero modes in equation
(4.4)) can be computed in a similar way. The result, that has to be added to
(4.15) is
$$
\eqalign{\cH_{-}|_{ker\cD^{\dagger}}(r,\theta;r',\theta')&=
{1\over 2\pi} e^{\int_{r'}^{r}
d\rho \at(\rho)/\rho}\sum_{i=0}^{[\Phi]} h_{i}(r,\theta;r',\theta')\cr
h_{i}(r,\theta;r',\theta')&=\left({r\over r'}\right)^{i+1/2}\times \left[ C^2
\int_{0}^{r}d\rho {e^{-2\int_{0}^{\rho} dx\at x^{-1}}\over \rho^{2i+1}} -
1\right]\cr}\eqno(4.16)
$$
where $C^2$ is a constant which makes zero the expression between brackets when
$r=R$.

For the computation of $\cH_{+}$ we follow the same steps, we expand the Green
function as in equation (4.8) in a basis of eigenfunctions of $\aR$ obtaining a
differential equation for the  coefficients of the expansion
$f_{\lambda}(r,r')$.
$$
\pr f_{\lambda}(r,r') + {1\over r}( n+{1\over 2}+\at(r))f_{\lambda}(r,r')=
\delta(r-r')\eqno(4.17)
$$
with the boundary conditions
$$
f_{\lambda}(R,r')=0~~~~~~~\if~~~R\lambda\leq -{1\over2}~~(n+1>-\Phi)
\eqno(4.18)
$$
$$
f_{\lambda}(r,R)=0~~~~~~~\if~~~R\lambda>-{1\over 2}~~(n+1\leq-\Phi).
\eqno(4.19)
$$
This equation has a solution very similar to the solution (4.13) and we find
for the Green function  $\cH_{+}$
$$
\cH_{+}(r,\theta;r',\theta')= e^{-\int_{r'}^{r}
d\rho \at(\rho)/\rho} \times \{\sum_{n+1\leq -\Phi} -{1\over
2}(1-sign(r-r'))\left({r'\over r}\right)
^{n+1/2} \nu_{\lambda}(\theta)\nu_{\lambda}^{*}(\theta')+
$$
$$
 +\sum_{n+1>
-\Phi}
{1\over 2}(1+sign(r-r'))\left({r'\over r}\right)^{n+1/2} \nu_{\lambda}(\theta)
\nu_{\lambda}^{*}(\theta')\}.\eqno(4.20)
$$
In terms of the holomorphic variables $z=re^{i\theta}$ and
$\omega=r'e^{i\theta'}$ the Green function can be written as
$$
\cH_{+}(r,\theta;r',\theta')={1\over 2\pi}
e^{-\int_{r'}^{r}
d\rho \at(\rho)/\rho}~\left({z\over \omega}\right)^{[\Phi]}
{\sqrt{z\omega}\over z-\omega}.\eqno(4.21)
$$

Then we have all the elements we need to compute fermionic correlation
functions.

As an application of the above results let us derive in our case the relation
between the chiral anomaly and the index formula. A gauge invariant definition
of the chiral current requires the introduction of an infinitesimal Wilson
line. Standard computation leads for the divergence of the chiral current
the value\refto{nielsen}
$$
\partial_{\mu}j_{\mu}^{5}=\lim_{\epsilon\to
0}{tr\left( \gamma_{\mu} \gamma^{5} \slD[A']^{-1}_{({\vec x} + {\vec
\epsilon}/2,{\vec x} - {\vec
\epsilon}/2)}\right)\epsilon_{\nu} F_{\mu \nu}[A']} + 4 \triangle \eta
- 2\sum_{zero~modes}
\varphi_{0 n} ({\vec x})~\varphi^{\dagger}_{0 n}({\vec y}).\eqno(4.22)
$$
Using the values (4.15), (4.16) and (4.21) for the Green function of the Dirac
operator we obtain, after taking the symmetric limit,
$$
\partial_{\mu}j_{\mu}^{5}={1\over 2\pi}\epsilon_{\mu \nu}F_{\mu \nu}[A'+a]
- 2\sum_{zero~modes}
\varphi_{0 n} ({\vec x})~\varphi^{\dagger}_{0 n}({\vec y}).\eqno(4.23)
$$
Then integrating over the whole disk we have
$$
\int_{\bf D}\partial_{\mu}j_{\mu}^{5}~d^2x=\oint_{\partial {\bf D}}
j_{\theta} d\theta= 2\Phi - 2~index~\slD.\eqno(4.24)
$$

Now we have to compute the boundary integral in (4.24). The current
$j_{\theta}$ at the boundary is given by
$$
j_{\theta}(R,\theta)=\lim_{\epsilon\to 0}{{1\over r^2}
\{\cH_{+}(R-\epsilon,\theta;R,\theta)+
\cH_{-}(R-\epsilon,\theta;R,\theta)\}}\eqno(4.25)
$$
and using equations (4.14) and (4.20) we see that this quantity tends to the
(ill defined) value
$$
{1\over 2\pi}\{\sum_{\lambda\leq -1/2} 1 - \sum_{\lambda>-1/2} 1\}.\eqno(4.26)
$$
But we just have found this quantity in equation (3.15) when we computed the
fermionic determinant. In fact this quantity, properly regularized, gives the
spectral asymmetry of the operator $\aR$. Using the results of section
III we find
$$
j_{\theta}(R,\theta) ={1\over \pi} (\Phi - [\Phi]).\eqno(4.27)
$$
Finally putting this result in equation (4.24) we recover again the index
formula
$$
index \slD=[\Phi].\eqno(4.28)
$$
We have to mention that this result, the relation between the anomaly equation
and the APS theorem, was first obtained by Niemi and Semenoff\refto{niemi} by a
different approach. Their computation, using the Pauli-Villars' regularization,
differs of that we presented here in the intermediate step (4.24) but both
procedures leads to the same, regularization independent, conclusion: the index
formula (4.28).

\head{\bf SECTION V}

In the precedent section we studied a Dirac action in a presence of a
background
gauge field with prescribed values at the boundary. Now let us consider the
case in which the gauge field is a quantum variable too (without kinetic term).
Examples of this kind of systems are the fermionic realization of the coset
models\refto{cosets}, the Thirring
model (in this case the coefficient in front of the bosonic action (3.18) is
different\refto{thirring}), \etc. Also we will allow the fermions to have
N colors. This last generalization is straightforward and only involves the
following changes: a sum over the color indices appears in equations (2.15) and
(2.16), the index theorem (2.30) takes the form
$$
index \slD = N[\Phi]\eqno(5.1)
$$
and the exponent of the jacobian in equations (3.18) and (3.26) is multiplied
by a factor $N$.

The quantization of the gauge field means, in the path-integral language, a
functional integral over the field the $A_{\mu}$ with the boundary conditions
(2.2). Instead of working
in the radial gauge $A_{r}=0$ it is more convenient to work in the gauge
$\omega=0$ (see equations (2.8) and (2.9)). (The value of the jacobian we
computed in section III (equation (3.18)) is gauge independent).
For simplicity let us choose a $\theta$-independent boundary condition for
the field $\ar$ \ie
$$
\aR(\theta)=const.={\Phi \over 2\pi R}\eqno(5.2)
$$
which in terms of the scalar field $\eta$ is the Neumann boundary
condition
$$
\pr \eta(r,\theta)|_{R}=0.\eqno(5.3)
$$

The bosonic action can be read from equation (3.18)
$$
{\cS}= -{1\over 2\pi}\int \eta \triangle \eta d^2 x + {\sqrt{N}\over \pi}\int
\eta F_{x y}[A'] d^2 x - (\Phi - [\Phi]){\sqrt{N}\over \pi}\oint \eta(R,\theta)
d\theta  + {1\over \sqrt{N}}\int \eta~ \j d^2 x\eqno(5.4)
$$
where we have included a source $\j$ to take into account the vertex operator
insertions $e^{\gamma^5 \eta}$ of equation (2.15) and we also rescaled $\eta\to
{1\over \sqrt{N}}\eta$.

The classical equations of motion are
$$
{1\over \sqrt{N}}\triangle \eta= F_{x y}-(\Phi-[\Phi]){\delta(r-R) \over R}
+ {\pi\over N}\j.\eqno(5.5)
$$
(Note that owing to the presence of the
delta function $\delta(r-R)$ the radial derivative of $\eta$ has a finite
discontinuity at the boundary).
Using the boundary condition (5.3) we can prove the identity
$$
\int_{\bf D} \triangle\eta({\vec x})d^2 x=0\eqno(5.6)
$$
which enforces the consistency condition
$$
\int_{\bf D} \j({\vec x})d^2 x=-2N[\Phi].\eqno(5.7)
$$
That is, the total charge created by the vertex operators must be
equal to $-2N[\Phi]$.

The solution of the equation of motion is obtained using the Green function of
the Laplacian
$$
\triangle G({\vec x},{\vec y})=\delta^2 ({\vec x},{\vec y}) \eqno(5.8)
$$ with
the boundary condition
$$
\pr G(r,\theta;r',\theta')|_{r=R}={1\over 2\pi R}\eqno(5.9)
$$
(the value $1/2\pi R$ in last equation is the only constant value consistent
with equation (5.8)).
This Green function can be computed with the help of the Poisson kernel and
gives
$$
G({r,\theta;r',\theta'})={1\over 4\pi} \log
\left[(r^2+r'^2-2rr'\cos(\theta-\theta'))(R^2+{r^2r'^2\over
R^2}-2rr'\cos(\theta-\theta'))/R^4\right].\eqno(5.10)
$$
Hence, using the second Green identity, we can write the solution of equation
(5.5) as
$$
\eta(r,\theta)=\sqrt{N}\int_{\bf D}d\theta' r'dr'G(r,\theta;r',\theta')\left(
F_{x y}(r',\theta') + {\pi\over N}\j (r',\theta')\right)+
$$
$$
-\sqrt{N}(\Phi-[\Phi])\left(\log R-H(r-R)\log({r\over R})\right)+
{1\over 2\pi}\oint_{\partial \bf
D}d\theta\eta(R,\theta)\eqno(5.11)
$$
where
$$
H(r-R)=\cases {0&if $r<R$\cr 1&if $r\geq R$\cr}\eqno(5.12)
$$
is the step Heaviside function. The last undetermined term in equation (5.11)
comes from the zero modes of the Laplacian (the Laplacian on the disk with the
Neumann boundary condition (5.3) has one zero mode which is a constant).

Now we can expand the field $\eta$ in the functional integral in modes of the
Laplacian operator around the classical solution (5.11) (the Laplacian
with the boundary condition (5.3) is self-adjoint). The integral over
the zero mode gives the charge balance condition (5.7)
$$
\int_{-\infty}^{\infty} d\eta_{0} e^{- \eta_{0}(\int_{\bf D}\j
d^2x-2N[\Phi])} = \delta\left(\int_{\bf D}\j d^2x-2N[\Phi]\right)\eqno(5.13)
$$
and the integrals over the remaining modes leads to the Gaussian term
$$
\sqrt{det'(\triangle)}~\exp \left(-{N\over 2\pi}\int d^2x d^2y (F_{x y}({\vec
x}) + {\pi\over N}\j ({\vec x}))G({\vec x},{\vec y})(F_{x y}({\vec x})
+{\pi\over N}\j ({\vec x}))\right)\eqno(5.14)
$$
with $G({\vec x},{\vec y})$ given by equation (5.10). Note that the
contributions of the boundary terms (terms containing $\delta(r-R)$) are absent
in this
result. One can prove that due to the
charge balance condition equation (5.7) these terms sums zero.

At this point we can compute fermionic correlation functions of the whole
theory by deriving the partition function (2.15) respect to the Grassmann
variables ${\bar \zeta}$ and $\zeta$ and then integrating the bosonic field
$\eta$. A rapid inspection of  equation (2.15) shows us that a necessary
condition for a non-zero value of the correlation function is that we must take
at least $N[\Phi]$ derivatives respect to the source ${\bar \zeta}_1$ and
$N[\Phi]$ derivatives respect to the source  $\zeta_1$. For the following
derivatives (if any) we must take the same number of derivatives respect to
${\bar \zeta}_1$ than $\zeta_2$, and the same number respect to  ${\bar
\zeta}_2$ than $\zeta_1$. Then a non-trivially zero v.e.v. has the form
$$
<0|\left({\bar \psi}_1\psi_1\right)^{N[\Phi]} \left({\bar \psi}_1\psi_2\right)^
q \left({\bar \psi}_2\psi_1\right)^p|0>\eqno(5.15)
$$
with $q$ and $p$ two arbitrary positive integers (the color indices and the
space-time variables are free). Note that after the chiral rotation (2.12), the
equation (5.15) is consistent with the charge balance condition (5.7).
The chiral rotation (2.12) decouple the bosonic excitations
from the fermionic ones and the following bosonic correlation
function factorizes of (5.15)
$$
F(x_1,...,y_{q+p})=<0|\prod_{i=0}^{2N[\Phi]+q+p}e^{\eta({\vec x}_{i})}~
\prod_{\alpha=0}^
{q+p}e^{-\eta({\vec y}_{\alpha})}|0>\eqno(5.16)
$$
(strictly speaking this result is valid if all the points $x_{i},y_{j},~
i=1,...,N[\Phi]+q+p;~j=1,...,q+p$ are different. If some of the points coincide
(\ie there are composite operators) we have to introduce some kind of
regularization which modifies the last expression as is the case of the chiral
current in section IV (equation (4.22)).

The quantity $F(x_1,...,y_{q+p})$ can be computed using equation (5.14) with
the particular election of the source \j
$$
\j ({\vec x})=\sum_{\alpha=1}^{q+p} \delta ({\vec x}-{\vec y}_{\alpha}) -
\sum_{i=1}^ {q+p+N[\Phi]} \delta ({\vec x}-{\vec x}_i).\eqno(5.17)
$$

Using the following result
$$
\oint G(r,\theta;r',\theta')d\theta=\oint G(r,\theta;r',\theta')d\theta'=
\cases {\log ({r\over R})&if $r>r'$\cr\log ({r'\over R})&if
$r<r'$\cr}\eqno(5.18)
$$
we can write for the terms in the exponential in (5.14) containing the magnetic
strength $F_{x y}$
$$
\int_{\bf D} F_{x y}({\vec x}) G({\vec x},{\vec y}) F_{x y}({\vec y}) d^2x
d^2y= -2\pi\int_0^R {\at^2(r)\over r} dr\eqno(5.19)
$$
$$
\int_{\bf D} F_{x y}({\vec x}) G({\vec x},{\vec y}) d^2x=
-\int_{|{\vec y}|}^R {\at(r)\over r} dr.\eqno(5.20)
$$

Hence the bosonic correlation function takes the form
$$
F=e^{N\int_0^R {\at^2(r)\over r} dr - \sum_i \int_{|{\vec x}_i|}^R
{\at(r)\over r} dr + \sum_\alpha \int_{|{\vec y}_\alpha|}^R {\at(r)\over r} dr}
$$
$$
e^{-{\pi\over 2N}\left(\sum_{i,j}G(x_i,x_j) + \sum_{\alpha,
\beta}G(y_{\alpha},y_{\beta}) - 2
\sum_{i,\alpha}G(x_i,y_{\alpha})\right)}\eqno(5.21)
$$
(the indices $i,j$ in the last equation run from 1 to $q+p+N[\Phi]$ and the
indices $\alpha,\beta$ from 1 to $q+p$).

For the second factor in equation (5.22) it is useful to define the variable
$$
x^*={R^2\over {\bar x}}~,~~~~{\bar x}= x_1 - i x_2\eqno(5.22)
$$
which is the inverse of $x=x_1 + i x_2$ through the circumference of radius R.
With this new variable the Green function (5.10) can be written as
$$
G(z,w)={1\over 4\pi}\log \left({|x-w||x^*-w^*||x-w^*||x^*-w|\over R^4}\right) +
{1\over 2\pi}\log {|z||w|\over R^2}.\eqno(5.23)
$$
In this form the effect of the geometry in the bosonic action (5.4) is the
appearance of a new vertex operator located at the ``image" point $z^*$ for
each vertex operator at the point $z$,  (see Cardy\refto{cardy}).

Because we are dealing with massless bosons the correlation function (5.16) has
u.v. divergencies manifested by the terms $G({\vec x},
{\vec x})$ of equation (5.21). We regularize this terms by introducing an
u.v. cut-off $a$ and defining the equal point Green function as
$$
G({\vec x},{\vec x})= {1\over 2\pi}\log\left({a\over R}\right) +
{1\over 2\pi}\log|x^* - x| + {1\over 2\pi}\log |x| .\eqno(5.24)
$$

Thus using equations (5.23) and (5.24) we can write the second factor of
$F(x_1,...,y_{q+p})$ as
$$
\prod_{i,j}|x_i-x_j|^{-{1\over 4N}}
\prod_{i,j}|x^*_i-x^*_j|^{-{1\over 4N}}
\prod_{i,j}|x^*_i-x_j|^{-{1\over 4N}}\times
$$
$$
\prod_{\alpha,\beta}|y_{\alpha}-y_{\beta}|^{-{1\over 4N}}
\prod_{\alpha,\beta}|y^*_{\alpha}-y^*_{\beta}|^{-{1\over 4N}}
\prod_{\alpha,\beta}|y^*_{\alpha}-y_{\beta}|^{-{1\over 4N}}\times
$$
$$
\prod_{i,\alpha}\left(|x_i-y_{\alpha}||x^*_i-y^*_{\alpha}||x^*_i-y_{\alpha}|
|x_i-y^*_{\alpha}|\right)^{1\over 4N}\times
$$
$$
\prod_{i}|x_i|^{1-2N[\Phi]\over 4N}\prod_{\alpha}|y_{\alpha}|
^{1+2N[\Phi]\over 4N}~a^{2(q+p)+N[\Phi]\over 4N},\eqno(5.25)
$$
where the products are taken over all the possible pairs of variables
without repetitions.

Now we have to compute the fermionic contribution to the correlation function
(5.15). Looking the generating functional (2.15) we easily verify that the
value of the fermionic correlator can be obtained following this rules: First
we chose $N[\Phi]$ fields ${\bar \psi}_1$ and $N[\Phi]$ fields  $\psi_1$ (one
of each color index) and replace each ${\bar \psi}_1$ by a different zero mode
$\varphi^{\dagger}_{n 1}$ and each $\psi_1$ by a different zero mode
$\varphi_{n 1}$ (with the same color indices). Then we sum over all the
different  possibilities of election of the zero modes. It easy to see that
this sum leads to the quantity
$$
\prod_{\mu=1}^N \left(\det \varphi_{i 1}(x_j^{\mu}) \times \det
\varphi^{\dagger}_{l 1}(x_m^{\mu})\right)\eqno(5.26)
$$
where
$x_j^{\mu},~~\mu=1,...,N;~j=1,...,[\Phi]$ are the coordinates of the fields
$\psi_1$ and $x_l^{\mu},~~\mu=1,...,N;~l=1,...,[\Phi]$ are the coordinates of
the fields ${\bar \psi}_1$ (in both cases $\mu$ is the color index).
Replacing the value of the zero modes by the ones of equation (4.7) we get for
(5.26)
$$
\prod_{\mu=0}^{N}\left(\prod_{j=0}^{2[\Phi]} {C_n\over \sqrt{2\pi}}
e^{-\int_0^{|x_j^{\mu}|} d\rho \at(\rho)/\rho}\right)\times
\prod_{\mu=0}^{N}
\left(\prod_{i<j}(z^{\mu}_i-z^{\mu}_j) \prod_{l<m}({\bar z}^{\mu}_l-{\bar z}^
{\mu}_m)\right).\eqno(5.27)
$$

With the remaining fermionic
fields we use the Wick theorem contracting all the possible pairs
$\underline{\psi_{1 j}
{\bar \psi}_{2 j}}$ and $\underline{\psi_{2 i} {\bar \psi}_{1 j}}$.
Then we replace each term $\underline{\psi_{1 j} {\bar \psi}_{2 j}}$ by
$-4\pi\sqrt{{\bar z}{\bar w}}\cH_-({\bar z},
{\bar w})\delta_{i j}$ and each term $\underline{\psi_{2 i}{\bar \psi}_{1 j}}$
by $-4\pi\sqrt{zw}\cH_+(z,w)\delta_{i j}$  ($i$ and $j$ are the color indices).
In particular we can verify that the exponential terms $e^{{+\atop
-}\int_{r'}^{r}  d\rho \at(\rho)/\rho}$ coming from $\cH_+$, $\cH_-$ (equations
(4.15) and (4.21)) and the determinant (5.27) cancel exactly the linear terms
in $\at$ in the exponential of the bosonic correlation function (5.21).

Finally we sum over all the different possibilities of election of the
term $\left({\bar \psi}_1 \psi_1\right)^{N[\Phi]}$. In all
the sums we have to take into account the sign due to the anticommuting
character of the Grassmann fields.

Multiplying this last result by the bosonic correlator (5.25) we obtain the
final answer for the fermionic correlation function function (5.15). The
generalization of this procedure to include composite fields (for example
current operators) is straightforward.

\head{\bf SECTION VI: CONCLUSIONS}

In this work we have carefully analysed different aspects of a gauged fermionic
theory defined on a disk. Because of the geometry special boundary conditions
({\it spectral} boundary conditions) have to be imposed to the fermions in
order to preserve charge conjugation and chiral invariance. These boundary
conditions guarantee the existence of $[\Phi]$ zero modes of the Dirac operator
($\Phi$, the magnetic flux).

Using the heat-kernel method we computed the functional determinant which
allowed us to map the arbitrary gauge field to a gauge field independent of the
angular variable with the same magnetic flux. We found that besides the usual
``anomaly density" terms, the determinant also contains a boundary term in
consistence with the APS theorem. In virtue of the $\theta$-independence  of
the gauge field we could compute exactly the Green functions of the resulting
fermionic theory. We also recovered the relation between the chiral anomaly and
the index theorem in an ``operational" way.

Finally we considered the coset model which results after the integration of
the gauge fields. The bosonic degrees of freedom of this constrained model can
be treated  as a theory of scalar bosons governed by a Liouville type action
with a conveniently chosen gravitational background field (functional of the
gauge field $\at'$). In this approach the geometry is manifested by the
presence of an interaction term with the geodesic curvature of the boundary.

Due to the nature of the scalar fields, originated from the gauge degrees of
freedom, they satisfy Neumann boundary conditions which affects deeply the
vacuum expectation values of vertex operators. For each vertex operator located
at the point $z$, a new ``image" vertex operator with the same charge appears
at the point $R^2/z$. We computed all the correlation functions of this theory.

\head{Acknowledgments} I would like to thank Eduardo Fradkin and Fidel
Schaposnik for useful discussions. This work was supported in part by the
National Science Foundation through grants No.DMR91-22385 and INT-8902032 at
the University of Illinois.

\references

\refis{solo} R. E. Gamboa Saravi, M. A. Muschietti and J. Solomin, \cmp 89,
363, 1983; \cmp 93, 407, 1984.

\refis{horda} M. Hortacsu, K. D. Rothe and B. Schroer, \np B171, 530, 1980;  M
Ninomiya and C. Tan, \np B257, 199, 1984;  Z. Ma, \journal J. Phys. {\bf A},
19,
L317, 1986.

\refis{brazil} S. A. Dias and M. T. Thomas, \prd 41, 1811, 1991.

\refis{aps} M. F. Atiyah, V. K. Patodi and I. M. Singer, \journal Math. Proc.
Camb. Phil. Soc., 77, 43, 1975.

\refis{determinantes} R. E. Gamboa Saravi, M. A. Muschietti, F. A. Schaposnik
and J. Solomin, \annp 157, 360, 1984.

\refis{rusos} Y. A. Sitenko, A. V. Mishchenko and Y. A. Sitenko,  preprint
ITP-91-16E.

\refis{mike} M. Stone, \annp 155, 56, 1984.

\refis{nielsen} N. K. Nielsen and B. Schroer, \np B127, 493, 1977.

\refis{niemi} A. J. Niemi and G. W. Semenoff, \np B269, 131, 1986.

\refis{yo} E. Moreno, to appear in {\it Mod. Phys. Lett.}{\bf A};
D. Cabra and E. Moreno, \journal J. Phys. {\bf A}, 23, 4711, 1990.

\refis{cosets} A. Polyakov, Lectures at Les Houches Session XLIX 1988, eds. E.
Brezin and J. Zinn Justin, Elsevier, 1989; K. Bardakci and C. Crescimanno, \np
B313, 69, 1989;  D. Cabra, E. Moreno and C. von Reichenbach, \journal Int.
Jour. of Mod. Phys. {\bf A} , 5, 2313, 1990.

\refis{thirring} K. Furuya, R. E. Gamboa Saravi and  F. A. Schaposnik ,\np
B208, 159, 1982.

\refis{cardy} J. L. Cardy, \np B240, 514, 1984.

\endreferences

\endit

\end